\pdfoutput=1

\documentclass[pra,10pt,superscriptaddress, footnoteinbib]{revtex4}
\usepackage{amsmath}
\usepackage{latexsym}
\usepackage{amssymb}
\usepackage{graphics,epstopdf}
\usepackage[colorlinks=true, citecolor=blue, urlcolor=blue ]{hyperref}
\usepackage{epsf,graphics,graphicx}

\newcommand{\ed}{\end{document}}
\newcommand{\beq}{\begin{equation}}
\newcommand{\eeq}{\end{equation}}

\begin{document}
	
\title	{ Quantum Capacitance of a Topological Insulator-Ferromagnet Interface} 
\author{Zhuo Bin Siu}
\affiliation{ Computational Nanoelectronics and Nanodevices Laboratory, National University of Singapore, Singapore}
 \author{Debashree Chowdhury}
 \affiliation{ Department of Physics, ​Harish-Chandra Research institute, Chhatnag Road, Jhusi, Allahabad, U. P. 211019,India}\email{debashreephys@gmail.com,Present address: ​Department of Physics,
 	Ben-Gurion University,
 	Beer Sheva 84105,
 	Israel }
 \author{ Mansoor B.A. Jalil}
 \affiliation{ Computational Nanoelectronics and Nanodevices Laboratory, National University of Singapore, Singapore}
 \author{Banasri Basu}
 \affiliation{Physics and Applied Mathematics Unit, Indian Statistical Institute, Kolkata 700108, India}
	\begin{abstract}
	We study the quantum capacitance in a  topological insulator thin film system magnetized in the in-plane direction in the presence of an out-of-plane magnetic field and hexagonal warping. To first order, the modification  in quantum capacitance due to hexagonal warping compared to the clean case, where both the in-plane magnetization and hexagonal warping are absent, is always negative, and increases in magnitude monotonically with the energy difference from the charge neutrality point. In contrast, the change in the quantum capacitance due to in-plane magnetization oscillates with the energy in general, except when a certain relation between the inter-surface coupling, out of plane Zeeman energy splitting and magnetic field strength is satisfied. In this special case, the quantum capacitance remains unchanged by the in-plane magnetization for all energies. 
	\end{abstract}
	
\maketitle	
	\section*{Introduction}
	Topological insulators (TI) \cite{DFerm1,Dferm2,JPSJ82_102001} are a new class of materials with some unique features. Three-dimensional TIs are metallic at the surfaces with an insulating bulk. Their topological nature provides robust surface states, which can be perturbed only by magnetic impurities. These distinctive features of TIs are the gateway for many interesting phenomenon \cite{PRB78_195424,PRL102_146805, PRL100_096407}. In particular, the suppression of back-scattering  \cite{PRL109_066803, Nat460_1106}  and the strong spin-momentum locking make TIs especially attractive material candidates in the arena of spintronics \cite{SpinD1,SpinD2, NatPhy5_378, NatMat11_409}. 
	
	TI thin films are of great recent interest to physicists as they can provide a captivating new avenue to eliminate the unwanted bulk contributions. TI thin films differ from semi-infinite bulk TI slabs in that they possess both top and bottom surfaces which can couple to each other across the finite thickness \cite{PRB80_205401, PRB81_041307,PRB81_115407}. This leads to a number of interesting effects, for example, the topological phase transition \cite{Zuzin, Menon, Menon1,zhuobin}, giant magneto-conductance effects \cite{PRL105_057401, PRB82_161104}, the quantum spin Hall effect \cite{PRB81_041307} and possible excitonic super-fluidity \cite{SpinA1}. In recent years, various novel devices based on TI and ferromagnets (FM) have been proposed. The TI/FM interface \cite{Ar1606_03812} is believed to be a potentially very interesting system to investigate. For example, there is a theoretical study \cite{bur} on the transport of charge as well as spin in TIs with FM electrodes. The quantum anomalous Hall effect \cite{Hybd1} and inverse spin-galvanic effect \cite{I} have also been investigated in this system. Besides, there have been numerous experimental studies on  the TI / FM interface \cite{B, Nat511_449, PRL114_257202}.  
	
	Furthermore, the Fermi surfaces of 3D TIs show some peculiarities, as their shapes are modulated by the electrochemical potential. The Hamiltonian of the 2D surface states of a 3D TI can be written as \cite{Fu} $H_{\text{2D}} = v_{F}(\sigma_{x}p_{y} - \sigma_{y}p_{x}),$ which exhibits $C_{3}$  symmetry as the Fermi velocity is the same in all in-plane directions. Thus the Fermi surface associated with $H_{\text{2D}}$ is circular. However, in ARPES experiments, the measured Fermi surface of some 3D TI materials deviate from the circular profile predicted by $H_{\text{2D}}$ at intermediate values of Fermi energy \cite{N,N1,N2}. Based on the symmetries of the underlying crystal structure, a cubic momentum correction factor, termed as hexagonal warping, has been proposed \cite{Fu} to explain the experimentally measured Fermi surfaces of some TIs like $\text{Bi}_2\text{Te}_3$.  As the electrochemical potential increases, one first  encounters a hexagonal Fermi surface which transforms to a snowflake geometry  at higher values of  electrochemical potential.
	
\begin{figure}
		\includegraphics[scale=0.6]{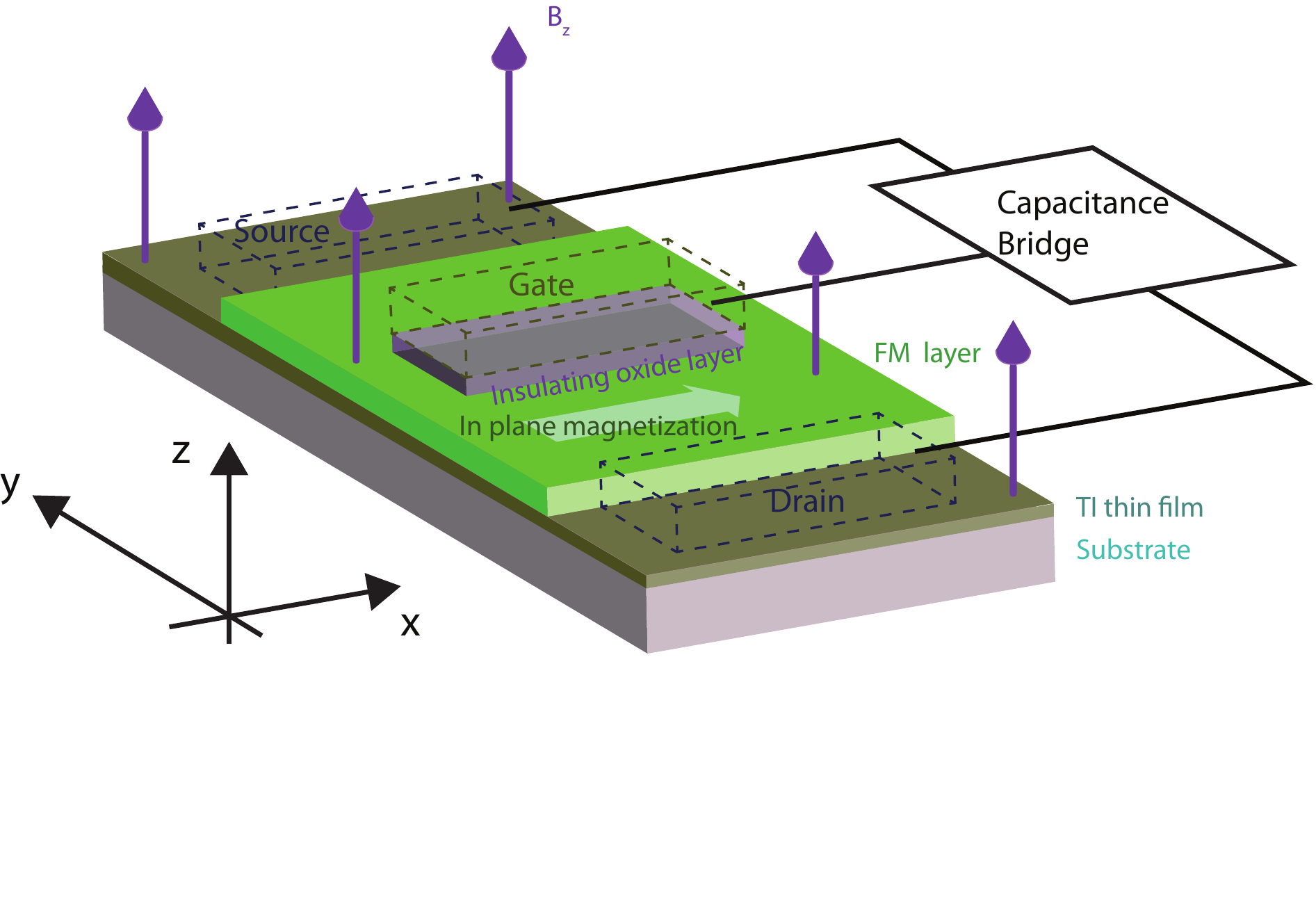}
		\centering
		\caption{A schematic diagram of the system under investigation consisting of a TI thin film under an adjoining FM layer with an in-plane magnetization subjected to an out-of-plane magnetic field.  The dotted boxes represent the source, drain and gate terminals in a typical quantum capacitance experimental setup connected to a capacitance bridge \cite{QC}.  } 
		\label{gmagSch} 
	\end{figure}

	On the other hand, capacitance in electrical devices is a well-known concept. In these systems there exists another contribution in the capacitance, known as quantum capacitance which is associated with the density of states (DOS) of the material \cite{SciRep4_5062}. For two-dimensional materials such as graphene, 2D electron gas  and the surface states of 3D TIs, quantum capacitance gives the dominant contribution in the capacitance measurement. A special feature of quantum capacitance is that it exhibits oscillations as the external magnetic field is varied \cite{Menon,T}. While the quantum capacitance in TI thin films in out-of-plane magnetic field with \cite{Menon} and without \cite{SciRep3_1261} the hexagonal warping has previously been studied, the effects of an in-plane magnetization \cite{m} on a TI thin film system with an out-of-plane magnetic field have, to the best of our knowledge, not been studied yet.  In the absence of the out-of-plane magnetic field, the in-plane magnetization lifts the energy degeneracy between the states localized near the top, and those near the bottom, of the film \cite{ Ar1606_03812}. The magnetization also leads to a $k$-space displacement of the two distinct Dirac cones corresponding to the states localized near the two surfaces even when hexagonal warping is present \cite{AIPAdv6_055706}. It is very natural to expect that the application of the  magnetic field will further modify the band structure and modulate the  DOS of the system. We are thus motivated to investigate  the effect of the in-plane magnetization on the oscillating pattern of the quantum capacitance for a TI thin film system with an in-plane magnetization and an out-of-plane magnetic field as schematically illustrated in Fig. \ref{gmagSch}. We have included the source, drain and gate terminals connected to a capacitance bridge in a typical measurement setup \cite{QC,CNT} in the figure. We refer the reader to the cited references for more details on the experimental measurement of the quantum capacitance --  in particular, Ref. \cite{CNT} has a schematic diagram for the quantum capacitance bridge circuit illustrated in the figure. 
	
	\section*{The ultrathin TI Hamiltonian and energy corrections}
	
	We write our model Hamiltonian as
	\beq H =  v_{F}(\sigma_{x}\pi_{y} - \sigma_{y}\pi_{x})\tau_z +\Delta \sigma_z + \Delta_t \tau_x + \frac{\lambda}{2}\sigma_{z}(\pi_{+}^{3} + \pi_{-}^{3}) + \vec{M} .\vec{\sigma}.
	\label{4}\eeq 
	The first term on the right hand side is due to spin-momentum locking effect, with $v_{F}$ being the Fermi velocity. $\sigma_{i}$ and $\tau_{i}$  with $i = x, y, z$ are the spin and the surface pseudospin degrees of freedom respectively. Here $\pi$ denotes the minimal momentum and $\tau_{z} =\pm 1$ denotes the states localized near the top / bottom surfaces  respectively. Compared to the Hamiltonian for a semi-infinite slab, the first term has an additional $\tau_z$ term with differing signs for the top and lower surfaces as the former has a normal in the $+\hat{z}$ direction and the latter in the opposite direction. The second term is due to the out-of-plane external magnetic field (B), with $\Delta =\mathfrak{r}B$.  $\mathfrak{r} = \frac{g}{2}\mu_{B}$, $g$ is the effective $g$-factor, $\mu_{B} = \frac{e\hbar}{2mc}$ is the Bohr magneton, $m$ is the free-electron mass, and $e$ is the electron charge.  Here we consider the magnetic field to be in the out-of-plane ($\hat{z}$) direction. $\Delta_{t}$ denotes the hybridization energy, which accounts for the tunnelling between the top and bottom surface states.  The fourth term $H_{p} = \frac{\lambda}{2}\sigma_{z}(\pi_{+}^{3} + \pi_{-}^{3}),$  is the hexagonal warping term, with $\lambda$ as the warping parameter. The last term is the in-plane magnetization term which arises due to the FM layer on the top of the TI thin film. A local axially-symmetric single-particle coupling appears in the system due to the ferromagnetic layer, which can be written as
	${
		H'=M(\cos(\phi_m)\sigma_x + \sin(\phi_m)\sigma_y)
		\label{2}}$
	where $M$ is the magnitude of the in-plane magnetization pointing in the $(M_x,M_y) = M(\cos(\phi_m),\sin(\phi_m))$ direction (we have absorbed the various coupling constants into the definition of $\vec{M}$).
	Thus one can rewrite Eq. (1) as, 
	$H = H_0 + H' + H_{p}$ where $H_0 = v_{F}(\sigma_{x}\pi_{y} - \sigma_{y}\pi_{x})\tau_z +\Delta \sigma_z + \Delta_t \tau_x .$	
	
	We now introduce the following ladder operators  
	$a = \frac{l}{\sqrt{2}}(\pi_{x} - i\pi_{y}),~~a^{\dag} = \frac{l}{\sqrt{2}}(\pi_{x} + i\pi_{y}),$ with $l = \sqrt{\frac{c}{eB}}$ as the magnetic length, in order to diagonalize $H$. Here $\vec{\pi} = -i\vec{\nabla} + \vec{A},$ is the gauge invariant momentum in the presence of the external magnetic field. Thus one can rewrite $H_0$ in terms of the ladder operators as \cite{Zuzin,Menon}
	\beq\label{5}
	H_0 = \frac{i \omega_c}{\sqrt{2}} \tau_z (\sigma_+ a - \sigma_- a^\dag) + \Delta \sigma_z + \Delta_t \tau_x,\eeq
	where $\omega_c = v_F / \ell$ is the characteristic frequency analogous to the cyclotron frequency of a usual 2DEG. 
	\begin{figure*}[ht!]
		\centering
		\includegraphics[scale=0.80]{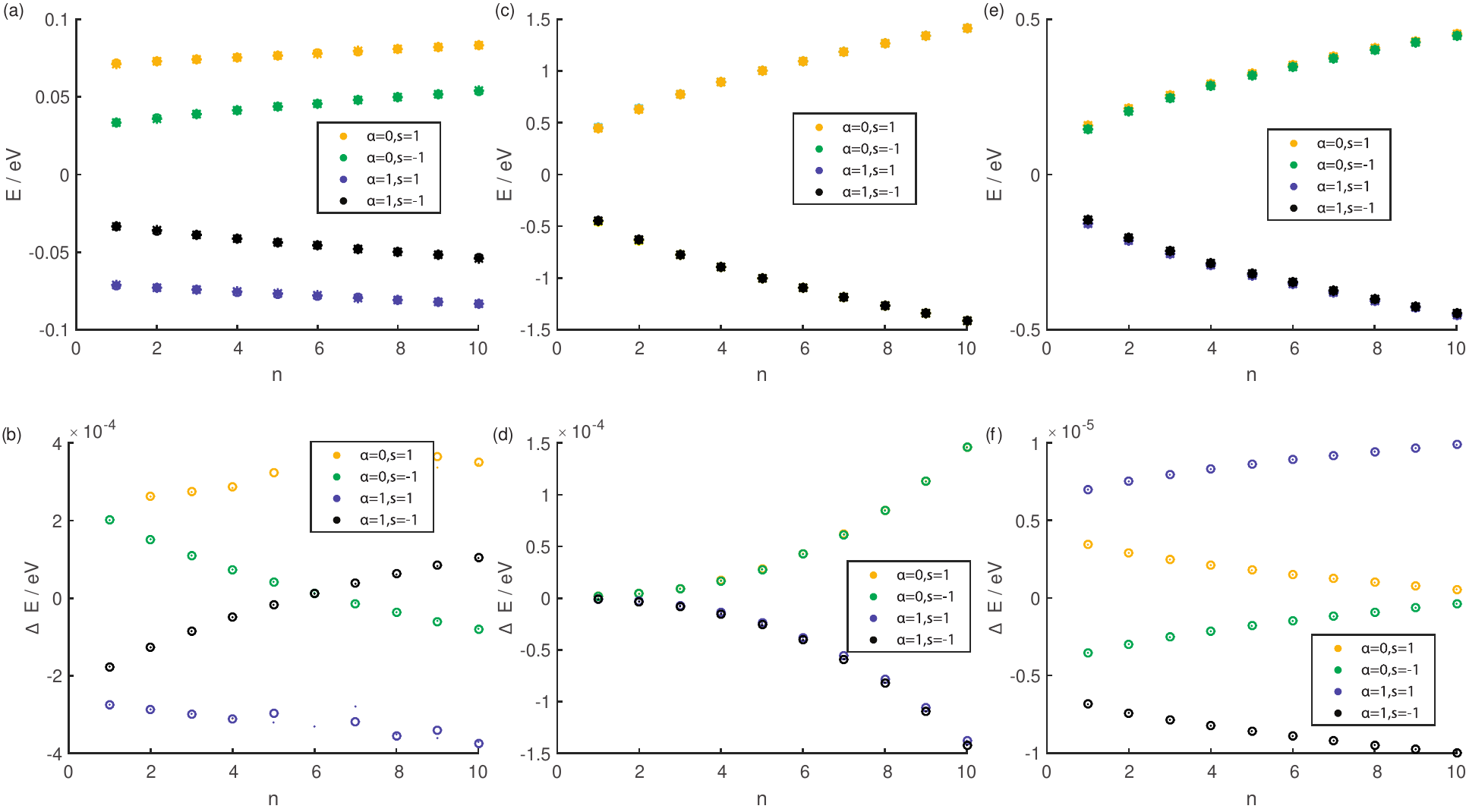}
		\caption{  The asterisks in the top panels  ( (a),(c),(e) ) depict the energy spectrum of the low energy states in a  TI thin film system where both the hexagonal warping and in-plane magnetization are absent. The open circles in these panels indicate the exact energy spectrum calculated (a) when only in-plane magnetization is present, (b) only hexagonal warping is present and (c) both hexagonal warping and in-plane magnetizations are present.  The lower panels ( (b), (d), (f) ) depict the exact energy corrections (open circles) and approximate second-order energy corrections (dots) for the respective panels above them.  ( The parameters $(\Delta t, \Delta z, \lambda, M, B)$ are for (a)/(b) $(0.02,0.05,0,0.005,10^{-4})$, (c)/(d) $(0.02,0.05,0.01,0,0.1)$ and (e)/(f) $(0.05,0.02,0.005,0.001,0.01)$ with units of eV for $\Delta t \Delta z$ and $M$,  $\mathrm{eV}\mathrm{nm}^3$ for $\lambda$ and T for $B$. )     } 
		\label{gEnCorr} 
	\end{figure*}	
	
	We can readily write the eigenstates  of the Hamiltonian $H_0$ as \cite{Zuzin,Menon}
	\begin{widetext} 
		\begin{eqnarray}
		\label{6}
		|n\alpha s\rangle = |n-1,\uparrow,T\rangle (-i s (-1)^\alpha f_{n\alpha s+}) + |n-1,\uparrow, B \rangle (i (-1)^{\alpha+s} f_{n\alpha s+} ) + |n,\downarrow, T \rangle(-s f_{n\alpha s-})+|n,\downarrow,B\rangle f_{n\alpha s-},
		\end{eqnarray}
	\end{widetext}
	where $| n, T(B), \uparrow(\downarrow) \rangle$ is the $n$-th Landau Level (LL) eigenstate pertaining to the Hamiltonian $H_{0}$ in Eq. \ref{5} localized near the top (bottom) surface with spin up (down). Here 
	$\alpha=0, 1$, $s = \pm$, $n = 0,\ldots,\infty$ and
	\beq
	\label{8}
	f_{n \alpha s \pm} = \frac{1}{2} \sqrt{1 \pm \frac{\Delta + s \Delta_t}{\epsilon_{n \alpha s}}}.
	\eeq
	Similarly one can readily obtain the energy eigenvalues of Hamiltonian $H_0$ as
	\beq
	\label{9}
	E_{n \alpha s} = (-1)^{\alpha}\sqrt{2 {\omega^2_{c}} n +({\Delta} + s{\Delta}_t)^2}.
	\eeq
	
	Now returning back to the Hamiltonian in Eq. \ref{4}, we have also the extra terms due to the warping and in-plane magnetization. We consider each of these terms separately. The an-harmonicity of the former and the breaking of the $z$ angular momentum conservation by the latter make it very difficult to obtain simple analytic expressions for the eigensystem for this Hamiltonian. We will hence employ the time independent perturbation theory to determine the corresponding energy corrections. 
	
	We consider the hexagonal warping term first. The first-order correction to energy can be obtained as (the superscript $(p)$ indicates that this energy correction is due to the hexagonal warping )
	$ E^{1(p)}_{n \alpha s} = \langle n\alpha s|\frac{\lambda}{2}\sigma_{z}(\pi_{+}^{3} + \pi_{-}^{3})|n\alpha s\rangle = 0.$ 
	Since the first-order energy correction is zero, we consider the second-order energy correction  \[ E^{2(p)}_{n \alpha s} = \sum_{n^{'} \alpha^{'} s^{'} \neq n \alpha s}\frac{|\langle n\alpha s|H_{p}|n^{'}\alpha^{'}s^{'}\rangle|^{2}}{E_{n} - E_{n^{'}}}. \] (Note that the summations in the above expression operate only on the primed variables. )
	After some algebraic manipulation, we have
	\begin{widetext}
		\begin{equation} 
		\begin{split}
		E^{2(p)}_{n \alpha s} =\frac{\lambda^{2}}{4l^{6}}\sum_{\alpha^{'}}\left\{\frac{(n-2)(n-1)}{E_{n\alpha s} - E_{n-3\alpha^{'}s}}\left[\sqrt{n-3}(-1)^{\alpha+\alpha^{'}}(1+\frac{\Delta +s\Delta_{t}}{E_{n\alpha s}})^{\frac{1}{2}}(1+\frac{\Delta +s\Delta_{t}}{E_{n-3\alpha^{'} s}})^{\frac{1}{2}}\right. \right.\\ \left.\left. - \sqrt{n}(1-\frac{\Delta +s\Delta_{t}}{E_{n\alpha s}})^{\frac{1}{2}}(1-\frac{\Delta +s\Delta_{t}}{E_{n-3\alpha^{'} s}})^{\frac{1}{2}}\right]^2+ \frac{(n+2)(n+1)}{E_{n\alpha s} - E_{n+3\alpha^{'}s}}\left[\sqrt{n}(-1)^{\alpha+\alpha^{'}}(1+\frac{\Delta +s\Delta_{t}}{E_{n\alpha s}})^{\frac{1}{2}}(1+\frac{\Delta +s\Delta_{t}}{E_{n+3\alpha^{'} s}})^{\frac{1}{2}}\right. \right.\\ \left.\left. - \sqrt{n+3}(1-\frac{\Delta +s\Delta_{t}}{E_{n\alpha s}})^{\frac{1}{2}}(1-\frac{\Delta +s\Delta_{t}}{E_{n+3\alpha^{'} s}})^{\frac{1}{2}}\right]^2\right\}.
		\end{split}
		\end{equation}
	\end{widetext}
	We now consider the in-plane magnetization. Rewriting $\vec{M}.\vec{\sigma} = M(\cos(\phi_m)\sigma_x + \sin(\phi_m)\sigma_y)$, we have, after some algebraic manipulations, 
	\begin{eqnarray}
	\langle n'\alpha' s| \vec{M}\cdot\vec{\sigma} |n,\alpha, s\rangle &=& \delta_{n',n\pm 1}\delta_{s',-s} \frac{M}{4}i^{1 + (1\mp 1)\alpha + (1 \pm 1)(\alpha'-s)} (-1+(-1)^s)  \nonumber \\
	&& \times \exp(\mp i \phi_m) \sqrt{1 \mp \frac{\Delta + s \Delta t}{e_{n,\alpha,s}}}\sqrt{1 \pm \frac{\Delta - s \Delta t}{e_{n \pm 1,\alpha',-s}}}. \label{vmv} 
	\end{eqnarray}
	
	The first-order energy correction due to $\vec{M}\cdot\vec{\sigma}$ vanishes while the second-order energy correction (the superscript $(m)$ denotes energy correction due to magnetization) is 
	\begin{widetext}
		\begin{eqnarray*}
			E^{2(m)}_{n\alpha s}= && \frac{M^2}{4 E_{n\alpha s}}  \sum_{\alpha'}\big\{ \\
			&& \frac{ (-\Delta + s\Delta_t + E_{n-1,\alpha',-s})(\Delta + s \Delta_t + E_{n,\alpha,s})}{E_{n-1,\alpha',-s}}\frac{1}{E_{n\alpha s}-E_{n-1,\alpha',-s}} + \\ 
			&& \frac{ (-\Delta - s\Delta_t + E_{n,\alpha',s})(\Delta - s \Delta_t + E_{n+1,\alpha,-s})}{E_{n+1,\alpha',-s}}\frac{1}{E_{n\alpha s}-E_{n+1,\alpha',-s}} \big\}.
		\end{eqnarray*}
	\end{widetext}
	The complete energy spectrum $E^T_{n \alpha s}$  is therefore
	\beq E_{n \alpha s}^{T}= E_{n \alpha s} + E_{n \alpha s}^{2(p)} + E_{n \alpha s}^{2(m)}. \eeq
	
	The exact energy spectrum can be computed by numerically diagonalizing $H$ written as a numerical matrix in the $|n, T(B), \uparrow (\downarrow) \rangle$ basis. For our numerical calculations we work in units where the numerical value of $v_F = \hbar = e = 1$, and energy is measured in electron volts.  We show in Fig. \ref{gEnCorr} there are parameter ranges where our analytic second-order energy corrections  reproduce the exact energy corrections reasonably well when either one or both hexagonal warping and in-plane magnetization are present.  In general, the magnitude of the energy correction due to the in-plane magnetization decreases with increasing magnitude of the applied magnetic field, while that for the hexagonal warping increases with the magnitude of the field. 
	
	We note in passing that our expressions for the energy correction due to in-plane magnetization do not reproduce the exact energy correction very well when the magnetization is of comparable or larger magnitude than the inter-surface coupling $\Delta_t$. This is because a quantum phase transition occurs when $|M| \geq |\Delta_t|$ \cite{Zuzin}, beyond which a perturbative approach to calculating the magnetization energy correction fails.\\

	\begin{figure*}[ht!]
		\centering
		\includegraphics[scale=0.8]{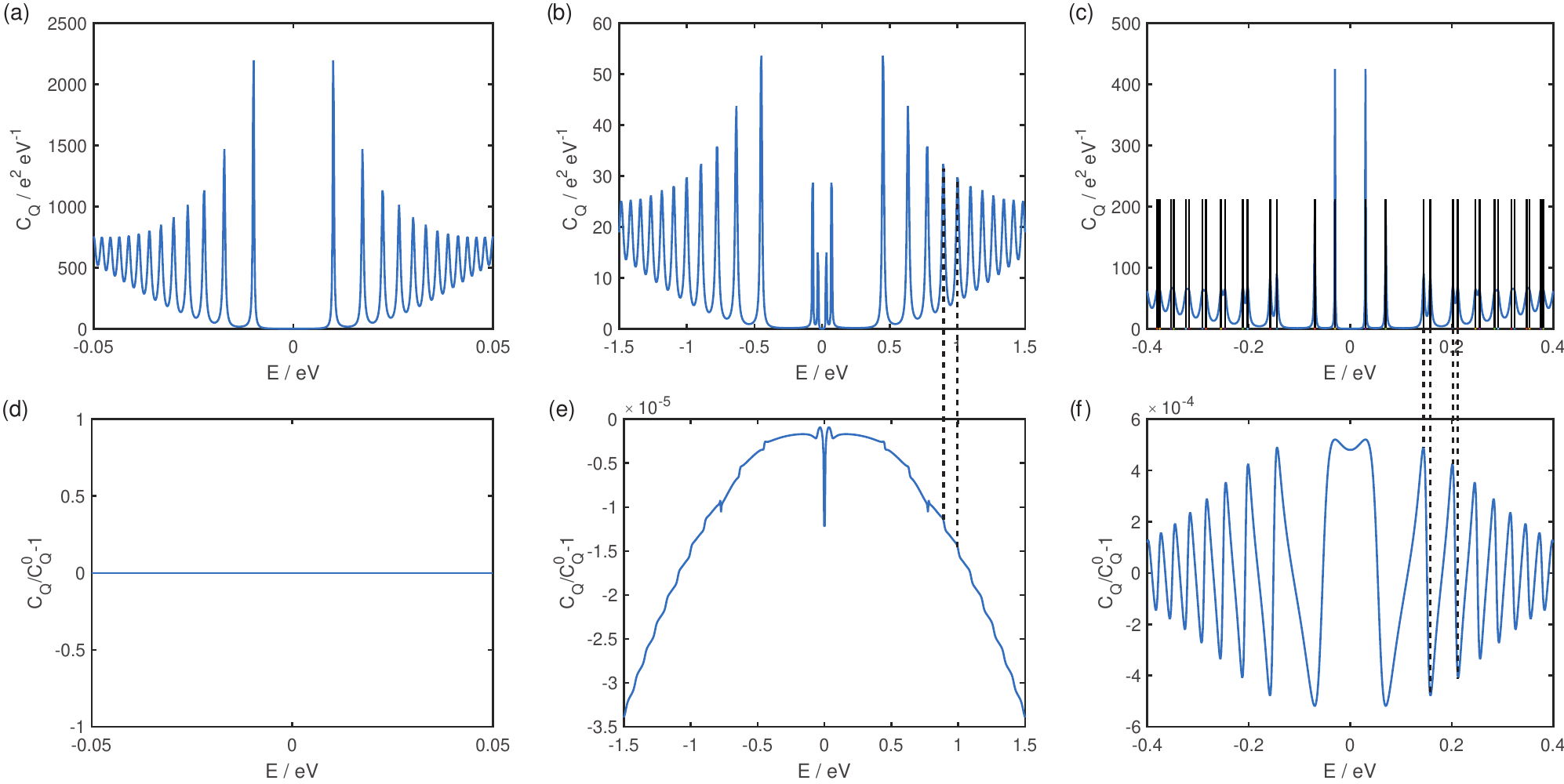}
		\caption{The top row ( (a) to (c) ) shows the quantum capacitance $C_Q$ for the same set of parameters as in Fig. \ref{gEnCorr} where (a) only in-plane magnetization is present, (b) only hexagonal warping is present and (c) both hexagonal warping and in-plane magnetization are present. The horizontal black lines in (c) indicate the energy values assumed by the discrete $|n,\alpha, s\rangle$ states.  Explicitly, the parameters $(\Delta t, \Delta z, \lambda, M, B, \Gamma_0)$ are for (a)  $(0.02,0.05,0,0.005,10^{-4},0.001)$ (b) $(0.02,0.05,0.01,0,0.1,0.03)$ and (c) $(0.05,0.02,0.005,0.001,0.01,0.01)$. The bottom row ( (d) to (e) ) shows the change in the quantum capacitance $(C_Q/C_Q^0-1)$ of their corresponding upper rows relative to their respective clean quantum capacitance $C_Q^0$ in the absence of both in-plane magnetization and hexagonal warping. The dotted lines joining panels (b) and (e), and (c) and (f), emphasise that features mentioned in the text in these panels emerge at the same energies. } 
		\label{gDOScomb} 
	\end{figure*}

	\section*{Quantum capacitance in magnetized TI}
	
 	There are many works reporting quantum capacitance measurements in carbon nanotubes and mono- and bi-layer graphene systems \cite{CNT, QC, Gra2, Gra3, Gra4} , which can form the basis for high-performance field effect transistors \cite{Gra5, Gra6}. In this present analysis, we are interested in the quantum capacitance in an ultra-thin TI/FM system with hexagonal warping and/or in-plane magnetization.  Before we start, let us first define what a quantum capacitor is. The usual classical capacitor formula for a parallel plate capacitor cannot be used when one plate of a parallel plate capacitor  has a low DOS.  Instead, we need to consider another contribution of the capacitor in series to the electrical capacitor. This extra contribution is the measure of the DOS at the Fermi energy and can be defined as \cite{T,1,2,3,QC} 
		\beq C_{Q} =e\frac{\partial Q}{\partial E}= e\frac{\partial^{2} n}{(\partial E)^{2}},\eeq
		 where $Q$ is the charge density due to electrons and $n$ is the carrier concentration at the Fermi energy \cite{T,1,2,3}.
		
	To compute the quantum capacitance, we need to first calculate the  DOS $D(E)$ in presence of the magnetic field and magnetization (see Supplementary Information ) at the Fermi energy.  It is to be noted here that the Fermi energy is the reference energy for the measurement of the quantum capacitance as we need to find out the shift of the Fermi energy due to the transfer of charge from the metallic conductor to the low  DOS TI material. If the TI is introduced as one of the electrodes of the capacitor, the total capacitance $C$ is significantly changed due to the electronic compressibility. The extra piece of the capacitance $C_Q$ is a direct measure of the DOS at the Fermi energy $E_{F}.$ 
	
	Quantum capacitance ($C_{Q}$) is readily defined as 
	\beq C_{Q} = e\frac{\partial Q}{\partial E_{F}} = e^{2}D_{T},\eeq where $E_{F}$ the Fermi energy and $D_{T}$ is the temperature dependent DOS, 
	\beq D_{T} = \int_{0}^{\infty}dE\frac{\partial f(E-E_{F})}{\partial E_{F}} D(E) .\eeq 
	
	Here, $f(E)$ is the Fermi-Dirac distribution function. In the limit of zero temperature, $D_{T=0} = D(E_{F})$, since the Fermi distribution approaches a Heaviside step function.  
	Substituting the expressions for values $E_{n,s}^{(2)}$ (see Supplementary Information for details), we obtain
	\begin{widetext}
		\begin{eqnarray}
		C_Q &=& \frac{e^2 D_0(E_f)}{2} \sum_s \left[ \frac{\exp(2(\frac{\pi \Gamma_0 E_F}{\omega_c^2})^2 )\cos(\pi \frac{[{E}^2 - (\Delta + s \Delta_t)^2]}{\omega_c^2})-1}{1 - 2\exp\{(\frac{\pi \Gamma_0 E_F}{ \omega_c^2})^2\}\cos(\pi \frac{[{E_F}^2 - (\Delta + s \Delta_t)^2]}{\omega_c^2}) + \exp\{2(\frac{\pi \Gamma_0 E_F}{\omega_c^2})^2\}}\right.  \sum_{\alpha}\Big[ 1 - \Big( \nonumber \\
		&&\frac{\lambda^{2}}{4l^{6}}\sum_{\alpha^{'}}\left\{\frac{(n-2)(n-1)}{E_{n_{0,s}\alpha s} - E_{n_{0,s}-3\alpha^{'}s}}\left[\sqrt{n-3}(-1)^{\alpha+\alpha^{'}}(1+\frac{\Delta +s\Delta_{t}}{E_{n_{0,s}\alpha s}})^{\frac{1}{2}}(1+\frac{\Delta +s\Delta_{t}}{E_{n_{0,s}-3\alpha^{'} s}})^{\frac{1}{2}}\right. \right. \nonumber \\ 
		&&\left.\left. - \sqrt{n}(1-\frac{\Delta +s\Delta_{t}}{E_{n_{0,s}\alpha s}})^{\frac{1}{2}}(1-\frac{\Delta +s\Delta_{t}}{E_{n_{0,s}-3\alpha^{'} s}})^{\frac{1}{2}}\right]^2+ \frac{(n+2)(n+1)}{E_{n_{0,s}\alpha s} - E_{n_{0,s}+3\alpha^{'}s}}\left[\sqrt{n}(-1)^{\alpha+\alpha^{'}}(1+\frac{\Delta +s\Delta_{t}}{E_{n_{0,s}\alpha s}})^{\frac{1}{2}}(1+\frac{\Delta +s\Delta_{t}}{E_{n_{0,s}+3\alpha^{'} s}})^{\frac{1}{2}}\right. \right. \nonumber  \\
		&& \left.\left. - \sqrt{n+3}(1-\frac{\Delta +s\Delta_{t}}{E_{n_{0,s}\alpha s}})^{\frac{1}{2}}(1-\frac{\Delta +s\Delta_{t}}{E_{n_{0,s}+3\alpha^{'} s}})^{\frac{1}{2}}\right]^2\right\}
		+\frac{M^2}{4 E_{n_{0,s}\alpha s}}  \sum_{\alpha'} \left\{ \right.  \frac{ (-\Delta + s\Delta_t + E_{n_{0,s}-1,\alpha',-s})(\Delta + s \Delta_t + E_{n_{0,s},\alpha,s})}{E_{n_{0,s}-1,\alpha',-s}} \nonumber \\ 
		&&\left. \frac{1}{E_{n_{0,s}\alpha s}-E_{n_{0,s}-1,\alpha',-s}} +  \left. \frac{ (-\Delta - s\Delta_t + E_{n_{0,s},\alpha',s})(\Delta - s \Delta_t + E_{n_{0,s}+1,\alpha',-s})}{E_{n_{0,s}+1,\alpha',-s}}\frac{1}{E_{n_{0,s}\alpha s}-E_{n_{0,s}+1,\alpha',-s}}\right\} \Big) \right]. \label{CqEq} 
		\end{eqnarray}
	\end{widetext}
	
	This final expression of quantum capacitance shows the dependence on the magnetization term and also on the warping parameter.  We stress that the expression above is derived by analytically summing over an \textit{infinte} numbers of states, and is derived by substituting our expression for the second-order correction to the energy $E_{n,s}^{(2)}$  as an approximation to the exact eigenenergies in the presence of the in-plane magnetization and / or hexagonal warping. We showed in Fig. \ref{gEnCorr} that the second order expression for the energy is a good approximation to the numerically calculated exact eigenenergies.

	We show in Fig. \ref{gDOScomb} the quantum capacitance $C_Q(E)$ at exemplary values of $\Gamma_0$ for the three sets of parameters as in Fig. \ref{gEnCorr} as well as the relative change in the quantum capacitance compared to the clean system $C^0_Q(E)$ where both the hexagonal warping and in-plane magnetization are absent. (The difference between $C_Q(E)$ and $C^0_Q(E)$ are not large enough for the difference between their plots to be visibly evident at the scale of the plots. ) Panel (c) shows that the peaks in the quantum capacitance occur at the discrete energy values of the Landau level states as expected. The pairs of closely-spaced Landau levels visible in panel (c) correspond to pairs of $|n,\alpha,s\rangle$ states with the same $n$ and $\alpha$ indices but different values of $s=\pm 1$. At small values of energy, the energy splitting between these pairs of states is large enough for the quantum capacitance to exhibit two separate peaks for $s=\pm 1$ for a given $n$ and $\alpha$, but as the energy magnitude increases the two peaks merge into a single broad peak.  
	
	Some features in the relative change in the quantum capacitance can be related to the DOS peaks which in turn correspond to the discrete energies of the $|n,\alpha,s\rangle$ states. Panel (e) shows that the hexagonal warping leads to a decrease in the quantum capacitance compared to the clean system. The extent of this decrease increases with energy, and step jumps occur at the energy values where new Landau states emerge. Although both hexagonal warping and in-plane magnetization are present in panel (f), the relative quantum capacitance change there is dominated by the in-plane magnetization. Unlike the hexagonal warping which leads only to a monotonic decrease in the quantum capacitance, the emergence of states with opposite signs of $s$ have opposite effects on the quantum capacitance. This leads to the oscillation of the quantum capacitance change in panel (f) where the local magnitude peaks in the quantum capacitance change correspond to the energies of the $|n,\alpha,s\rangle$ eigenstates. 
	
	The effects of the hexagonal warping has already been studied at some length in an earlier work by some of the authors \cite{Menon}. We focus here instead on the effect of the in-plane magnetization on the quantum capacitance. The change in the quantum capacitance due to the in-plane magnetization happens to be 0 for the parameters in panel (a) of Fig. \ref{gDOScomb}. The quantum capacitance change due to the magnetization is, in general, finite as we will show shortly. Let us first explain why the quantum capacitance change is zero for this particular set of parameters. 
	
	The quantum capacitance can be written in the form of (see the Supplementary Information),  
	\begin{equation}
	C_Q(E) = e^2\frac{{D_0}(E)}{2} \sum_{\alpha, s} c(\alpha,s) \left(1-\frac{ E^{(2)}_{n_{0,s}\alpha s}}{\omega_c^2}\right)  \label{DE1} 
	\end{equation}
	where
	\begin{eqnarray*}
		&& c(\alpha,s) = \\
		&& \frac{\exp(2(\frac{\pi \Gamma_0 E}{\omega_c^2})^2 )\cos(\pi \frac{[{E}^2 - (\Delta + s \Delta_t)^2]}{\omega_c^2})-1}{1 - 2\exp\{(\frac{\pi \Gamma_0 E}{ \omega_c^2})^2\}\cos(\pi \frac{[{E}^2 - (\Delta + s \Delta_t)^2]}{\omega_c^2}) + \exp\{2(\frac{\pi \Gamma_0 E}{\omega_c^2})^2\}}. 
	\end{eqnarray*}
	Notice that $s$ appears in $c(\alpha,s)$ only inside the $\cos(\pi \frac{[{E}^2 - (\Delta + s \Delta_t)^2]}{\omega_c^2})$ terms. When the difference between $\frac{[{E}^2 - (\Delta + s \Delta_t)^2]}{\omega_c^2}$ with $s=1$ and $s=-1$ is an integer,  i.e. 
	\begin{equation}
	\frac{ 2\Delta\Delta_t }{\omega_c^2} \in \mathbb{Z} \label{noMagD},
	\end{equation}
	we have $c(\alpha,s=1) = c(\alpha, s=-1)$ so that the $c(\alpha,s)$ term can be factorized out to give 
	\begin{eqnarray}
	&&(C_Q  - C_Q^0)(E) e^2\Big |_{\frac{2\Delta\Delta_t}{\omega_c^2}\in \mathbb{Z}} \nonumber \\ 
	&=&  -\frac{C_Q^0(E)}{2\omega_c^2}  \sum_{\alpha} c(\alpha, s=1) \Big( \nonumber  E^{(2)}_{n_{0,s=+1},\alpha,s=+1} + E^{(2)}_{n_{0,s=-1},\alpha,s=-1}    \Big) \label{noMagD2} 
	\end{eqnarray}
	The two terms on the last line $( E^{(2)}_{n_{0,s=+1},\alpha,s=+1} + E^{(2)}_{n_{0,s=-1},\alpha,s=-1} )$ have the same magnitude but opposite signs (even when Eq. \ref{noMagD} is not satisfied) so they cancel each other out and Eq. \ref{noMagD2} evaluates to zero. The set of parameters for Fig. \ref{gDOScomb}(a) satisfies Eq. \ref{noMagD} and hence the quantum capacitance change is zero. 
	
\begin{figure}[ht!]
		\centering
		\includegraphics[scale=0.6]{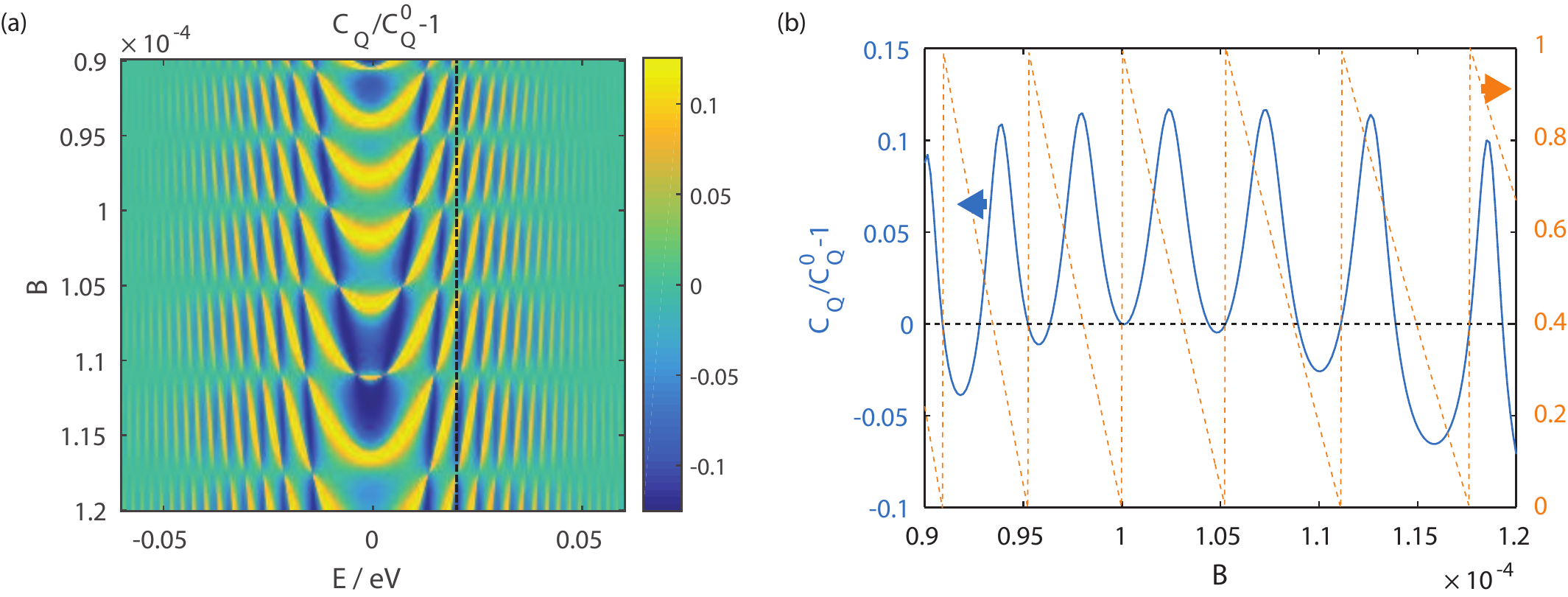}
		\caption{ Panel (a) shows the relative quantum capacitance change due to the in-plane magnetization as a function of the energy and the magnetic field with the same set of parameters as Fig. \ref{gDOScomb}(a) except for the magnitude of the magnetic field. Panel (b) shows the relative quantum capacitance change at $E=0.02\ \mathrm{eV}$ (indicated by the dotted line in panel (a) ) as a function of the magnetic field (the left axis), and (right axis) the fractional part of $2\Delta\Delta_t/\omega_c^2$, $\mathrm{mod}(\frac{2\Delta\Delta t}{\omega_c^2}, 1)$ where $\mathrm{mod}(a,b)$ is the modulus of $a$ divided by $b$. }
		\label{gDOSRatComb} 
	\end{figure}

	On the other hand when Eq. \ref{noMagD} is not satisfied Eq. \ref{DE1} can no longer be factorized as per Eq. \ref{noMagD2} because $c(\alpha,s=1) \neq c(\alpha, s=-1)$, and we do obtain a finite quantum capacitance change. This is illustrated in Fig. \ref{gDOSRatComb} which shows the relative quantum capacitance change for the same parameter set as in Fig. \ref{gDOScomb}(a)/(d) except for the magnitude of the magnetic field. Most of the features in panel (a) of the figure are qualitatively similar to that in panel (f) of Fig. \ref{gDOScomb}. A new feature evident in Fig. \ref{gDOSRatComb} is that the magnitude of the relative quantum capacitance change oscillates with the magnetic field. In particular, there are values of $B$ in which the quantum capacitance change is zero. Panel (b) of the figure shows that these values of $B$ correspond to those which satisfy Eq. \ref{noMagD} -- the vertical green lines, representing those values of $B$ where the $\mathrm{mod} ( \frac{2\Delta\Delta t}{\omega_c^2}, 1 )$ jumps to 0 when $\frac{2\Delta\Delta t}{\omega_c^2}$ is integer, coincides with the values of $B$ where the quantum capacitance change is zero. 
	
	\section*{Conclusion}
	In this paper we have considered the influence of the hexagonal warping and in-plane magnetization in a FM/TI thin film system with an out-of-plane magnetic field on the DOS of the system. The quantum capacitance, which is the measure of the DOS at Fermi energy shows some oscillations when plotted against the Fermi energy with the peaks corresponding to the discrete Landau levels. The hexagonal warping always leads to a decrease in the quantum capacitance with the magnitude of the decrease increasing with the energy difference from the charge neutrality point. By contrast the sign and magnitude of the quantum capacitance change due to the in-plane magnetization oscillate with the energy, and goes to zero for all energies when Eq. \ref{noMagD} is satisfied. \\
	
	\section*{Acknowledgments} 
	Z. B. S. and M. B. A. J.  acknowledge the financial support of MOE Tier II grant MOE2013-T2-2-125 (NUS Grant No. R-263-000-B10-112) and Tier I grant R-263-000-B98-112, and the National Research Foundation of Singapore under the CRP Programs ``Next Generation Spin Torque Memories: From Fundamental Physics to Applications'' NRF-CRP12-2013-01 and ``Non-Volatile Magnetic Logic and Memory Integrated Circuit Devices'' NRF-CRP9-2011-01.
	
	\section*{Author Contributions} 
	Z. B. S. and D. C. performed the calculations and wrote most of the manuscript text. M. B. A. J. and B. B. contributed to the discussion and improvement of the manuscript. 
	
	\section*{Additional Information}
	\subsection*{Competing financial interests} The authors declare no competing financial interests.

	\section{Supplementary information}
	\underline{Calculation of Density of states}\\
	
	The DOS can be expressed in terms of self energy as
	\beq D(E) = \mathrm{Im}[\frac{\Sigma^{-}(E)}{\pi^{2}l^{2}}\Gamma_{0}^{2}],\nonumber\eeq 
	
	where  $ \Sigma^{-}(E) = \Gamma_{0}^{2}\sum_{n} \sum_{\alpha, s} \frac{1}{E -E_{n\alpha s}^T-\Sigma^{-}(E)} = \sum_{\alpha, s} \Sigma^{-}_{\alpha, s} (E), \label{18}$ is the self energy of the system with $\Gamma_0$ as the impurity induced LL broadening. The modified energy of the system is $E_{n \alpha s}^{T} = E_{n \alpha s} + E_{n \alpha s}^{(2)},$ where $E^{(2)}_{n\alpha s} \equiv E^{2(p)}_{n\alpha s} + E^{2(m)}_{n\alpha s}$. As $E_{n}^{(2)} << E_{n}$ (we omit the indices $\alpha$ and $s$ when no confusion might arise), we have, for each value of $\alpha$ and $s$,   
	\\

	\beq \Sigma^{-}_{\alpha, s} (E) = \Gamma_{0}^{2}\sum_{n}\frac{1}{E -E_{n,\alpha,s}-\Sigma^{-}(E)} + \Gamma_{0}^{2}\sum_{n}\frac{E_{n\alpha s}^{(2)}}{[E -E_{n,\alpha,s}-\Sigma^{-}(E)]^{2}},\nonumber\eeq 
	
	on Taylor expansion. The first term is easy to calculate and we finally can obtain  
	\beq \Gamma_{0}^{2}\sum_{n}\frac{1}{E -E_{n \alpha s}-\Sigma^{-}(E)} =  \frac{\pi \Gamma_{0}^{2} E}{w_{c}^{2}} \cot(\pi n_{0, s})\nonumber\eeq
	where 
	$n_{0, s} = \frac{1}{(2\hbar \omega_{c}^{2})}[\{E - (\Delta +s\Delta_{t})^{2}]$ is the pole neglecting the contribution of the self-energy. Note that $n_{0,s}$ has the same value regardless of whether $\alpha = 0$ or 1 so we have omitted the $\alpha$ index in $n_{0s}$.  \\
	\\
	The second term in the self energy expression contains a repeated pole and using standard summation formula we have, temporarily dropping the $\alpha$ and $s$ indices for notational brevity, 
	
	\beq	\Sigma^-(E)_{\alpha,s \text{2nd term}} = -\frac{\pi\Gamma_0^2} {\omega_c^4}\Big(\omega_c^2 \cot(\pi n_0) E^{(2)}_{n_0} - \pi\csc(\pi n_0)^2 (E_{n_0})^2 E^{(2)}_{n_0} + \cot(\pi n_0) (E_{n_0})^2 (\partial_z E^{(2)}_{n_0})\Big) 
	\eeq

	Noting that $E\ll 1,$ neglecting higher order terms in $E_{n_0}$ , lands us at 
	
	
	\beq 
	\Sigma^{-}_{\alpha, s} (E) = \frac{\pi \Gamma_0^2 E}{\omega_c^2} \left(1-\frac{ E_{n_{0, s}\alpha s}^{(2)}}{\omega_{c}^{2}}\right)\cot \pi n_{0, s}.\eeq
	
	Within the $E^2_{n_0},$ by summing over each $s$ and $\alpha$ branch we can, following the method of Refs. \cite{JPhyC26_165303} and \cite{Menon}  extract the total DOS as 
	
	\beq
	D(E)  = \sum_{s=\pm 1, \alpha}  \frac{ D_0(E)}{2} \left[ 1 + 2\sum_{u=1}^\infty \exp\{-u(\frac{\pi\Gamma_{0}E}{\omega_{c}^2})^2\}\cos\left( u \frac{\pi}{\omega_c^2}(E^2 - (\Delta + s \Delta_t)^2 \right)\right]\left(1-\frac{ E^{(2)}_{n_{0,s}\alpha s}}{\omega_{c}^{2}}\right)
	\eeq
	where $D_0(E) = \frac{2 E}{\pi \omega_c^2}$, $n_{0,s} = \frac{1}{\omega_{c}^{2}}\left[E^{2} - (\Delta +s \Delta_{t})^{2}\right]$. 
	We've recovered the asymptotic form ($n>1$) of the DOS with a minute additional term which manifests due to warping and in-plane magnetization. 
	
	One can further rewrite the DOS by expressing the summation on the right hand side as  $\mathrm{Re}  \sum_{u =1, \infty} ( \exp(u(-\gamma^2 + i\beta))$ where $\gamma = (\frac{\pi\Gamma_{0}E}{\omega_{c}^2})$ and $\beta = \frac{\pi}{\omega_c^2}(E^2 - (\Delta + s \Delta_t)^2)$. Evaluating the sum gives 
	\[
	\mathrm{Re}  \sum_{u =1}^\infty ( \exp(u(-\gamma^2 + i\beta)) = \frac{ \exp(-\gamma^2)(\cos\beta - \exp(-\gamma^2))} { 1 -2\cos\beta\exp(-\gamma^2) + \exp(-2\gamma^2)}.
	\]
	The DOS can then be written as 
	\begin{equation}	
	D(E) = \sum_{s,\alpha} D(E, s, \alpha) \label{Desa0}
	\end{equation}
	where
	\beq 
	D(E,s, \alpha) = \frac{{D_0}(E)}{2}  \frac{\exp(2(\frac{\pi \Gamma_0 E}{\omega_c^2})^2 )\cos(\pi \frac{[{E}^2 - (\Delta + s \Delta_t)^2]}{\omega_c^2})-1}{1 - 2\exp\{(\frac{\pi \Gamma_0 E}{ \omega_c^2})^2\}\cos(\pi \frac{[{E}^2 - (\Delta + s \Delta_t)^2]}{\omega_c^2}) + \exp\{2(\frac{\pi \Gamma_0 E}{\omega_c^2})^2\}}  \left(1-\frac{ E^{(2)}_{n_{0,s}\alpha s}}{\omega_c^2}\right). \label{Desa}
	\eeq
\end{document}